\documentclass{article}

    \usepackage[nonatbib,final]{neurips_2020}

\usepackage[utf8]{inputenc} 
\usepackage[T1]{fontenc}    
\usepackage{hyperref}       
\usepackage{url}            
\usepackage{booktabs}       
\usepackage{amsfonts}       
\usepackage{nicefrac}       
\usepackage{microtype}      
\usepackage{amssymb,amsthm}
\usepackage[super,negative]{nth}
\usepackage{subcaption}
\usepackage{mdframed}
\usepackage{graphicx}
\usepackage{enumitem}
\usepackage{varwidth}
\usepackage{hyperref}
\usepackage[noend]{algpseudocode}
\usepackage{amsmath}
\usepackage{tabularx}
\usepackage{algorithm}
\usepackage{algorithmicx}
\usepackage{booktabs}
\usepackage{calc}
\usepackage{listings}
\usepackage{textcomp}
\usepackage{upquote,textcomp}
\usepackage{enumitem}
\usepackage{xfrac}
\usepackage{numprint}
\usepackage{mathtools}
\usepackage{float}
\usepackage{array}
\usepackage{xcolor, colortbl}
\usepackage{multicol}
\usepackage{varwidth}
\usepackage{dashrule}
\usepackage{enumitem}
\usepackage{bookmark}
\usepackage{todonotes}
\usepackage{xspace}
\usepackage{todonotes}
\usepackage{lipsum}
\usepackage{multirow}
\algrenewcommand\alglinenumber[1]{\tiny #1:}

\newcommand{\onesgx}{{1-TEE}\xspace}
\newcommand{\nsgx}{{N-TEE}\xspace}
\newcommand{\secureboost}{{SecureBoost}\xspace}

\title{Mitigating Leakage in Federated Learning\\ with Trusted Hardware}

\author{%
  Javad Ghareh Chamani\\
  Hong Kong University\\
  of Science \& Technology\\
  \texttt{jgc@cse.ust.hk} \\
  \And
  Dimitrios Papadopoulos \\
  Hong Kong University\\
  of Science \& Technology\\  \texttt{dipapado@cse.ust.hk} \\
}

\begin{document}

\maketitle

\begin{abstract}
In federated learning, multiple parties collaborate in order to train a global model over their respective datasets. Even though cryptographic primitives (e.g.,  homomorphic encryption) can help achieve data privacy in this setting, some partial information may still be leaked across parties if this is done non-judiciously.  In this work, we study the  federated learning framework of \secureboost [Cheng et al., FL@IJCAI'19] as a specific such example, demonstrate a leakage-abuse attack based on its leakage profile, and experimentally evaluate the effectiveness of our attack. We then propose two secure versions relying on \emph{trusted execution environments}. We implement and benchmark our protocols to demonstrate that they are 1.2-5.4$\times$ faster in computation and need 5-49$\times$ less communication than \secureboost.

\end{abstract}

\section{Introduction}\label{sec-introduction}
Federated machine learning (FML)~\cite{yang2019federated,konevcny2016federated,bonawitz2017practical} is increasingly used by companies and organizations that want to collaboratively train a model based on the union of their separate datasets  (e.g., banks and insurance companies storing their customers' data). Typically, in FML a local model is iteratively trained by each participant and is then merged into the global model. In this manner, it is not necessary to transfer raw data across collaborators; the exchanged messages necessary in order to compute the global model consist of local model parameters, gradients, partially trained models, etc.

One of the main motivations of FML is to maintain data privacy~\cite{yang2019federated}. This may be necessary due to  conflicts of interest in the market or legally mandated from privacy policies such as GDPR. To achieve data privacy,  collaborating parties should \emph{at the very least} not transfer raw data. However, this does not guarantee privacy on its own. The transferred values (e.g., model parameters, partial models, gradients) may still leak some information that can be used to infer raw data. A growing line of works on privacy-preserving machine learning (e.g.,~\cite{nikolaenko2013privacy,aono2016scalable,mohassel2018aby3,shokri2015privacy,privacy2018phong,DBLP:journals/corr/abs-1904-04475}) use cryptographic techniques such as secure multi-party computation and homomorphic encryption to ensure no information is revealed across parties, other than the (final) global model. However, using cryptographic techniques does not automatically guarantee privacy. When using such primitives in the context of a complex system, it is still possible that some partial information leaks which can be exploited by leakage-abuse attacks~\cite{grubbs2017leakage,xu2019constructive,kellaris2016generic,kornaropoulos2020state,lacharite2018improved,markatou2019full,seal}.

In this work, we study the partial leakage of \secureboost~\cite{cheng2019secureboost} (a lossless vertical federated tree-boosting system proposed by WeBank). \secureboost allows a federated-learning process to be executed by multiple parties with partially common data samples but different features. It is based on XGBoost~\cite{chen2016xgboost} and tries to predict the classification label via regression trees. In \secureboost, one participant (called \emph{active party}) holds sensitive labels that should not be revealed to the other participants (called \emph{passive} parties). On the other hand, passive parties do not want to reveal any information about their data samples to each other or to the active party. Although \secureboost employs additive homomorphic encryption during the tree node-splitting process, the way it is used reveals the \emph{partial ordering of passive parties' samples along each feature to the active party}. To evaluate the effect of this leakage, first we perform a leakage-abuse attack 
that tries to guess the data sample values. We evaluate our attack (for different prior knowledge assumptions) on a public financial dataset of 150K samples~\cite{dataset}. Under the (mild) assumption that the attacker knows some approximate distribution of the values (e.g., via public census data) we can guess values with very high accuracy (e.g., age and salary within $\pm$5 years and $\pm$2000\$, respectively, with more than 99\% accuracy). 

We then propose two modified versions of \secureboost that mitigate this leakage by relying on trusted execution environment (TEE) hardware~\cite{mckeen2013innovative,kaplan2016amd}. This follows a line of recent works~\cite{ohrimenko2016oblivious,hynes2018efficient,zhang2020enabling,mo2019efficient, tramer2018slalom} that propose using trusted hardware for privacy-preserving distributed learning. Our first solution (\nsgx) assumes all parties have access to TEE. It achieves  asymptotically optimal computation time and communication. Our second solution (\onesgx) makes a more light-weight assumption, i.e., it only assumes TEE is available at the active party. However, it is somewhat less efficient than our first one. We implemented both schemes and integrated them with the FATE~\cite{fate} FML framework. According to our experimental evaluation, our solutions not only eliminate the aforementioned leakage but are also 1.2-5.4$\times$ faster in computation time and require 5-49$\times$ less communication size than \secureboost.

We believe this work can help highlight the subtle issue of leaked information in federated learning, even when cryptographic primitives are (partially) used, and how TEEs can help address it. There are many future directions, such as improving the performance of our schemes and combining cryptographic primitives and trusted hardware to get secure and fast solutions for other FML tasks.
\vspace{-.1cm}

\section{Preliminaries}\label{sec-preliminaries}
\noindent\textbf{Trusted Execution Environment (TEE).}
Secure hardware, such as Intel-SGX~\cite{mckeen2013innovative} and AMD enclave~\cite{kaplan2016amd}, provides a trusted environment for users to execute their code in an untrusted setting, even when assuming that the operating system itself may be compromised. It provides three important properties: (i) \emph{Isolated execution} which is achieved by reserving a portion of the system's memory used to store the user's code and data in encrypted form. (ii) \emph{Sealing} which allows secure persistent storage of data on the untrusted area. (iii) \emph{Remote attestation} which ensures the correctness of the executed code in the enclave. We stress that our secure solutions can operate with any trusted hardware that realizes these properties. This is particularly important in view of the recent attacks against Intel-SGX~\cite{van2018foreshadow,lipp2018meltdown}.

\smallskip\noindent\textbf{\secureboost.} Cheng et al.~\cite{cheng2019secureboost} proposed \secureboost   as a lossless framework for gradient boosting decision trees on vertically-partitioned data. In this federated machine learning setting each party has its own features for a common set of data samples. At a high level, \secureboost operates as follows. First, it uses a privacy-preserving entity alignment technique \cite{liang2004privacy} to find the common samples among the parties' datasets. Then, it executes an iterative computation between the active party that has the sensitive classification labels and the passive parties who only have samples' values for their own features. This part is based on the XGBoost~\cite{chen2016xgboost} framework. In each iteration, the active party computes the gradients of samples belonging to the current tree nodes and sends their Paillier homomorphic encryption to all the passive parties. Due to the semantic security and homomorphism of Paillier encryption~\cite{paillier1999public}, this allows each passive party to compute all the local possible splits (across its samples and features), while protecting the sensitive gradient values. Hence, each passive party iterates over its own features and samples values’ thresholds in order to compute all possible local splits' gradient summations. Then, it sends all of them back to the active party. The active party decrypts the received encrypted splits from all parties, computes their corresponding objective function scores, and finds the best global split by comparing them. For completeness, the \secureboost procedures are provided in Appendix~\ref{sec:appendix}.

\section{Proposed Leakage Abuse Attack}\label{sec-attack}
As mentioned in the introduction, non-judicious use of cryptographic primitives can lead to leakage of information. In this section, we explain the data leakage profile of \secureboost and our proposed attack. In \secureboost, after the setup and entity alignment phase, in each round the active party computes the Paillier homomorphic encryption of samples' gradients (denoted by $g_i$ and $h_i$) and sends them to all passive parties, as explained above. The active party then \emph{decrypts} the sum of the gradients and computes each split's score to determine the best global split. Although the active party only needs to learn the value of the \emph{optimal} split, it actually learns \emph{all splits of all features for every party}. Clearly, this is a lot of additional information. Our main observation is that, based on this information, an (honest-but-curious) active party can correlate the provided sums of gradients in order to estimate the passive parties' samples' partial ordering across each feature. The experimental evaluation of our attack can be found in Section~\ref{sec-experiments-attack}.

As a concrete example, assume that there is one passive party with one feature and 3 samples \{$x_1=20,x_2=30,x_3=15$\}, and one active party with gradients \{$g_1=-1,g_2=0.6,g_3=0.2$\}. The information sent by the passive party to the active party is: \{$split_1=(\langle 0.2 \rangle,\langle -1 \rangle+\langle 0.6 \rangle), split_2=(\langle 0.2 \rangle+\langle -1 \rangle,\langle 0.6 \rangle)$\} where $\langle .\rangle$ denotes Paillier encryption. The active party decrypts the first split and gets (0.2,-0.4). Comparing this pair with possible sum combinations of $g_i$ values reveals that 0.2 is related to sample $x_3$ and -0.4 is the sum of the other two samples. Likewise, the second split reveals that $x_2$ is greater than or equal to the other two samples. Therefore, the active party can infer that $x_3\le x_1 \le x_2$ for this feature.

\section{Secure Protocols with TEE}\label{sec-solutions}
We now present our solutions for securing \secureboost, based on trusted hardware. Our protocols are modified versions of \secureboost that only change the split-finding procedure; other parts remain the same. The first one assumes that each party has access to a TEE (\nsgx) while the second one assumes that only the active party has a TEE (\onesgx).  At the beginning of both protocols, code is loaded into the enclaves and interested parties get a code attestation. Then, the active party's enclave generates a key for a semantically secure symmetric encryption scheme (e.g., AES) and a secure channel is established between it and the passive parties' enclaves (for \nsgx) or the passive parties directly (for \onesgx), e.g., using~\cite{baumann2015shielding,schuster2015vc3}. Secret keys are communicated (as necessary) via this secure channel. In what follows, we explain one iteration of the \secureboost with our modified schemes.



\smallskip\noindent\textbf{\nsgx.} The basic idea behind \nsgx is to first find the best local split of each passive party within its trusted hardware and then find the global best split within the active party's trusted hardware. The split-finding procedure of \nsgx is presented in Algorithm~\ref{alg-ntee}. First, the active party's enclave computes the symmetric encryption of gradients using the secret key for the aligned samples (denoted by $I$) and sends them to the passive parties' enclaves (line 1). Each passive party's enclave decrypts the gradients using the secret key (originally communicated by the active party's enclave) and iterates over all features and values' thresholds to find its best local split (lines 2-18). For simplicity, we assume that all encryptions can fit in the enclave memory. Otherwise, we have to use a paging mechanism for loading and unloading the ciphertexts. Then, each passive party's enclave encrypts its best local score and sends it to the active party's enclave (line 19). The active party's enclave decrypts all local best splits and determines the best global split by comparing their scores (lines 20-24). It finally returns this output to the active party.

\smallskip\noindent\textbf{1-TEE.} \onesgx works assuming just a single TEE at the active party. The process is presented in Algorithm~\ref{alg-1tee}. At a high level, the active party uses one-time pad for encryption of gradients ($g_i$ and $h_i$). This hides the actual gradient values from passive parties while providing additive homomorphism in a very efficient way (lines 1-2). Similar to standard \secureboost, passive parties compute all possible splits according to their local features and threshold values (without any TEE assistance). Then, they encrypt all splits and their corresponding used gradients' indexes using the secret key (originally communicated by the active party's enclave) and send them back to the active party (lines 3-7). The gradients' indexes are needed for consistently removing the one-time pad randomness from the provided split sums at the active party. When the active party receives the encrypted splits, it passes them along with the the random values used for one-time pad encryption to its enclave. Finally, the enclave decrypts all splits, removes their randomness based on the used gradients' indexes (lines 15,17), finds the best global split (lines 8-22), and returns it to the active party.

\smallskip\noindent\textbf{Security and Efficiency.} In both schemes the secret key is exchanged through a secure channel and a semantically secure symmetric encryption scheme is used for encryption. Values are accessed by the TEE through sequential scans hence accesses are data-independent and leak no information. Therefore, there is no leakage and passive parties learn nothing while the active party only learns the global best split outputted by its TEE. Regarding the performance, \nsgx and \onesgx replace the costly Paillier encryption of \secureboost with much more light-weight symmetric encryption which makes them  more efficient. Moreover, \nsgx has asymptotically optimal communication, much better than \secureboost, whereas \onesgx has asymptotically the same communication as \secureboost but it is concretely better since it uses one-time pad instead of Paillier. 

\vspace{-.2cm}

\begin{algorithm}[t]
    \centering
    \caption{N-TEE}\label{alg-ntee}
    \vspace{-.4cm}
    \begin{multicols}{2}
    \begin{algorithmic}[1]
        \item[Active Party Enclave]
            \State Send $Enc_{sk}(g_i)$ and $Enc_{sk}(h_i)$ for all $i\in I$ to all Parties
        \item[]
        \item[Each Passive Party Enclave]   
            \State $Dec_{sk}(g_i)$ and $Dec_{sk}(h_i)$
            \State $g\leftarrow \sum_{i\in I}g_i$, $h\leftarrow \sum_{i\in I}h_i$
            \State $best_{scor}\leftarrow-\infty$
            \State $best_f\leftarrow-1$ ; $best_{tr}\leftarrow -1$
            \State //enumerate all features
            \For{$k=0$ to $d$} 
                \State $g_l=0$ ; $h_l=0$
                \State //enumerate all threshold values
                \For{$v=0$ to $l_k$}    
                    \State $g_l \leftarrow g_l + \sum_{i\in\{i|s_{k,v}\ge x_{i,k}>s_{k,v-1}\}}g_i$
                    \State $h_l \leftarrow h_l + \sum_{i\in\{i|s_{k,v}\ge x_{i,k}>s_{k,v-1}\}}h_i$
                    \State $g_r \leftarrow g-g_l$, $h_r \leftarrow h-h_l$
                    \State $score = \frac{g_l^2}{h_l+\lambda}+\frac{g_r^2}{h_r+\lambda}+\frac{g^2}{h+\lambda}$
                    \If{$score > best_{scor}$}
                        \State $best_{scor} \leftarrow score$
                        \State $best_f\leftarrow k$
                        \State $best_{tr}\leftarrow v$
                    \EndIf
                \EndFor
            \EndFor
            \State Send $Enc_{sk}(best_{scor})$ to Active Party
        \item[]
        \item[Active Party Enclave]
            \State $best_{scr}\leftarrow -\infty$, $best_{indx}\leftarrow -1$
            \For{$i=0$ to $m$}  \Comment{\# of parties}
                \State $curScore$ = $Dec_{sk}($Party$_i.best_{scor})$
                \If{ $curScore > best_{scr}$}
                        \State $best_{scr} \leftarrow curScore$; $best_{indx}\leftarrow i$
                \EndIf
            \EndFor
            
    \end{algorithmic}
    \end{multicols}
    \vspace{-.3cm}
\end{algorithm}

\begin{algorithm}[t]
    \centering
    \caption{1-TEE}\label{alg-1tee}
    \vspace{-.4cm}
    \begin{multicols}{2}
    \begin{algorithmic}[1]
        \item[Active Party]
            \State $g_i^{rnd}\xleftarrow{\$}[0,\infty]$ , $h_i^{rnd}\xleftarrow{\$}[0,\infty]$ for all $i\in I$
            \State Send $g_i'\leftarrow g_i+g_i^{rnd}$ and $h_i'\leftarrow h_i+h_i^{rnd}$ for all $i\in I$ to all Parties
        \item[]
        \item[Each Passive Party]   
            \For{$k=0$ to $d$} \Comment{enumerate all features} 
                \State \hspace{-7pt} $G_{kv}\leftarrow Enc_{sk}(\sum_{i\in\{i|s_{k,v}\ge x_{i,k}>s_{k,v-1}\}}g_i')$
                \State \hspace{-7pt}$H_{kv}\leftarrow Enc_{sk}(\sum_{i\in\{i|s_{k,v}\ge x_{i,k}>s_{k,v-1}\}}h_i')$
                \State \hspace{-7pt}$I_{kv}\leftarrow Enc_{sk}(\{i|s_{k,v}\ge x_{i,k}>s_{k,v-1}\})$
            \EndFor
            \State Send $G_{kv}$, $H_{kv}$, and $I_{kv}$ to Active Party
        \item[]    
        \item[Active Party Enclave]   
            \State $g\leftarrow \sum_{i\in I}g_i$, $h\leftarrow \sum_{i\in I}h_i$, $best_i\leftarrow-1$
            \State $best_{score} \leftarrow -\infty$ , $best_f\leftarrow-1$
            \State $best_{tr}\leftarrow -1$
            \For{$i=0$ to $m$}\Comment{\# of parties}
                \For{$k=0$ to $d_i$}  \Comment{\# of features}
                    \State $g_l=0, h_l=0$
                    \For{$v=0$ to $l_k$}    \Comment{\# of thresholds}
                        \State $t\leftarrow Dec_{sk}(G_{kv}^i) -\sum_{i\in I_{kv}^i}g_i^{rnd}$
                        \State $g_l \leftarrow g_l + t$
                        \State $t\leftarrow Dec_{sk}(H_{kv}^i) -\sum_{i\in I_{kv}^i}h_i^{rnd}$
                        \State $h_l \leftarrow h_l + t$
                        \State $scr = \frac{g_l^2}{h_l+\lambda}+\frac{g_r^2}{h_r+\lambda}+\frac{g^2}{h+\lambda}$
                        \If{$scr > best_{score}$} 
                            \State $best_i\leftarrow i$; $best_{score} \leftarrow scr$
                            \State $best_f\leftarrow k$; $best_{tr}\leftarrow v$
                        \EndIf
                    \EndFor
                \EndFor
            \EndFor
    \end{algorithmic}
    \end{multicols}
    \vspace{-.3cm}
\end{algorithm}

\section{Experimental Evaluation}\label{sec-experiments}
In this section, we evaluate our leakage-abuse attack accuracy and measure the performance of our constructions. We implemented our schemes in C++ using Intel-SGX and integrated them with FATE~\cite{fate} (excluding the one-time setup of secure channels between enclaves). 
We conducted our experiments on a public financial dataset~\cite{dataset} (the one that was used in the SecureBoost paper~\cite{cheng2019secureboost}) with 150K samples and 10 attributes. All experiments were executed on a machine with a four-core Intel Xeon E-2174G 3.8GHz processor with SGX capabilities, Ubuntu18.04 LTS,  64GB RAM, 1TB SSD, and AES-NI enabled.

\subsection{Leakage-Abuse Attack}\label{sec-experiments-attack}
The first set of experiments demonstrates the effectiveness of our attack for two features: Age (between 21 and 109) and Salary (between 0\$ and 3M\$). For each feature, we measure the accuracy of the attacker's guess using four methods: (i) Random Min-Max: guess randomly knowing only the minimum and maximum values. (ii) Our Attack with Min-Max: first sort  samples based on the inferred partial order and then assign the values (knowing the minimum and maximum). (iii) Random with Approximate Distribution: guess randomly assuming knowledge of a 10\% increment approximate distribution~\cite{islam2012access,grubbs2017leakage}. (iv) Our Attack with Approximate Distribution: first sort samples based on the inferred partial order and then assign the values (knowing  a 10\% increment approximate distribution).

\begin{figure}[t!]
	\centering
	\begin{subfigure}[t]{0.24\linewidth}
		\includegraphics[width=\textwidth]{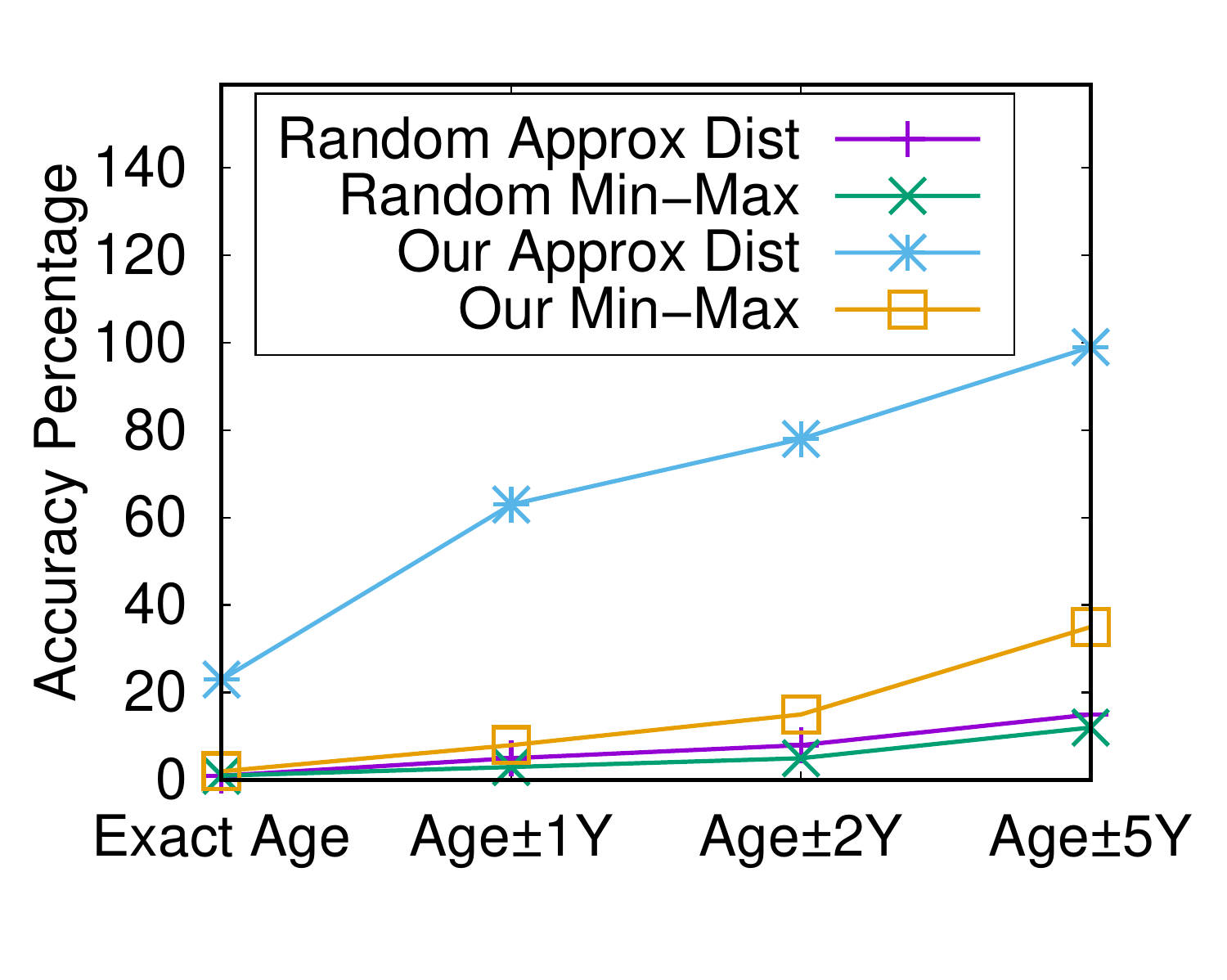}
		\vspace{-.6cm}
		\caption{}
	\end{subfigure}%
	\begin{subfigure}[t]{0.24\linewidth}
		\includegraphics[width=\textwidth]{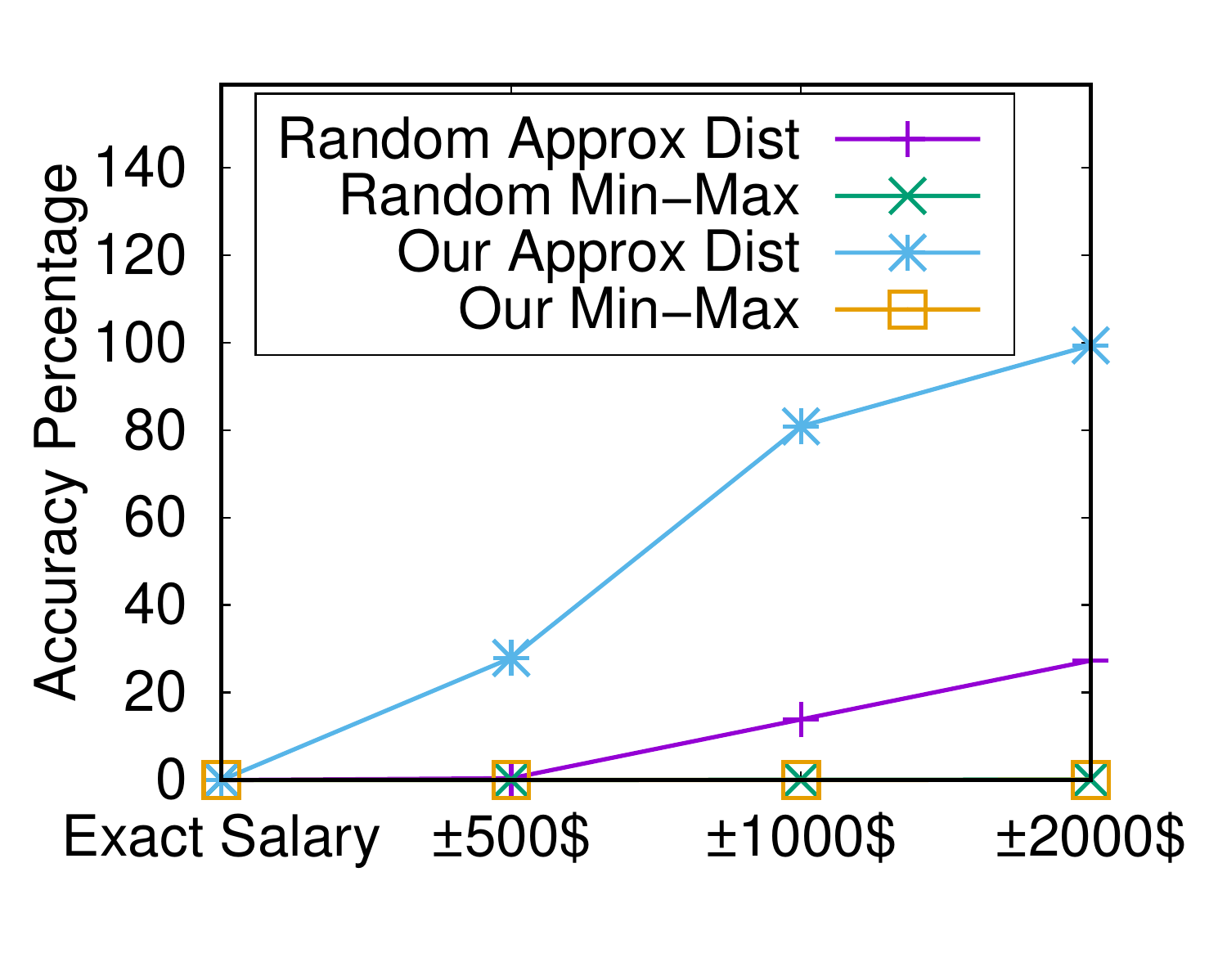}
		\vspace{-.6cm}
		\caption{}
	\end{subfigure}
	\begin{subfigure}[t]{0.24\linewidth}
		\includegraphics[width=\textwidth]{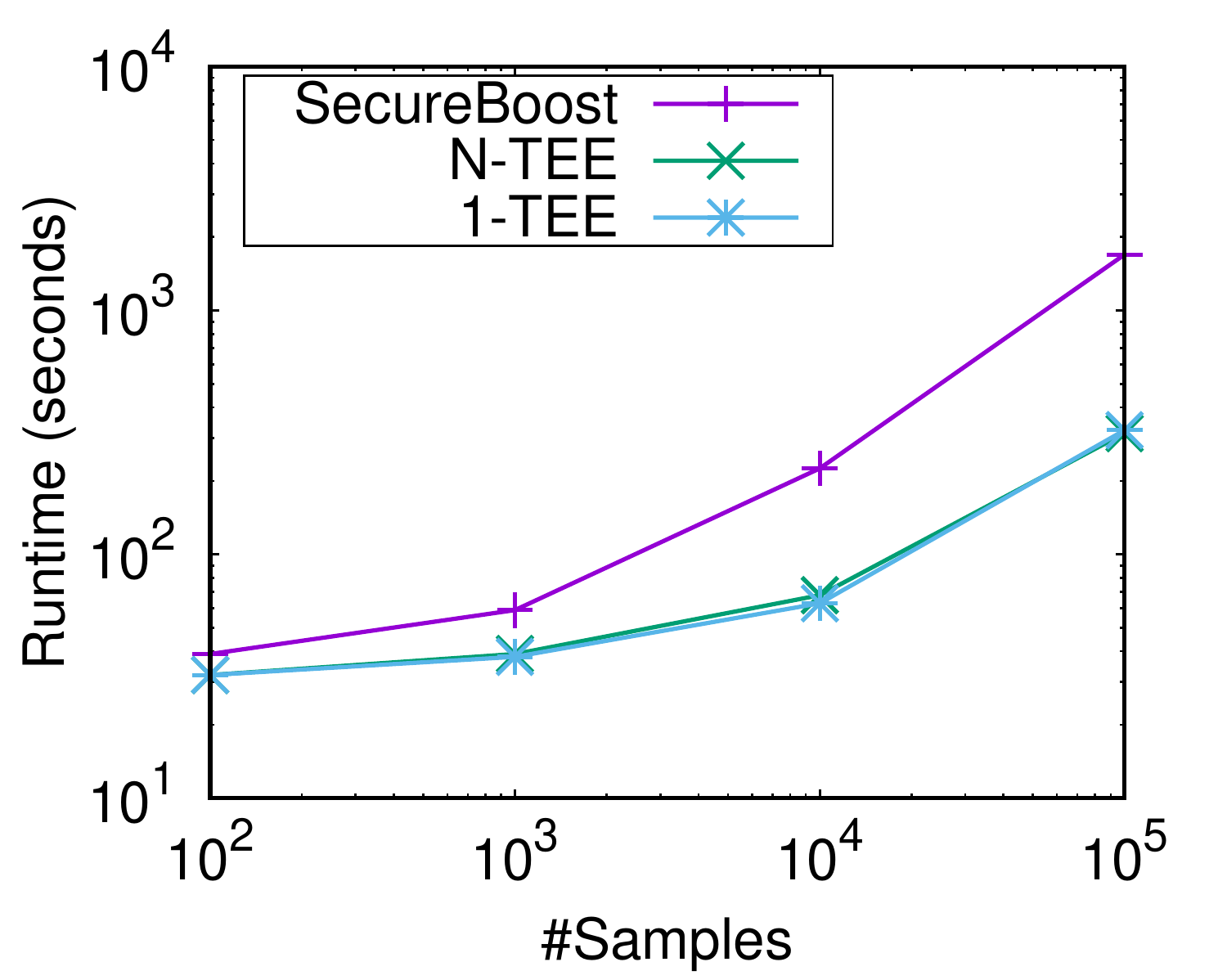}
		\vspace{-.6cm}
		\caption{}
	\end{subfigure}%
	\begin{subfigure}[t]{0.24\linewidth}
		\includegraphics[width=\textwidth]{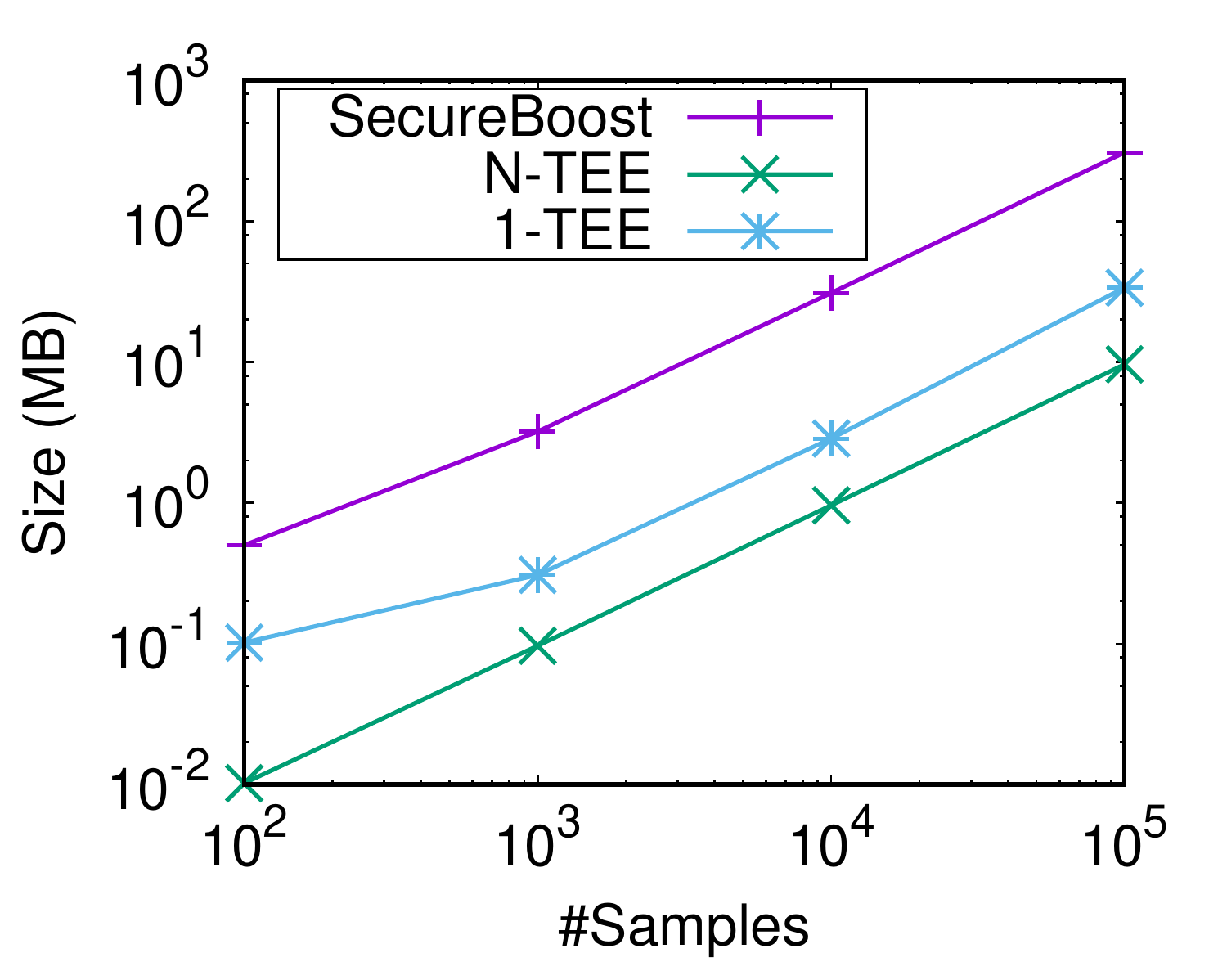}
		\vspace{-.6cm}
		\caption{}
	\end{subfigure}	\vspace{-.2cm}
	\caption{Our proposed attack accuracy for guessing (a) age, (b) salary. Our proposed secure solutions' evaluation for (c) computation time, and (d) communication size.}
	\label{fig:performance}
	\vspace{-.3cm}
\end{figure}

The accuracy of guessing the age and salary values for 150K samples using the above four methods is presented in Figures~\ref{fig:performance}(a),(b). The obvious conclusions from the figures are as follows: (i) our approximate-distribution-based attack has the best accuracy among all other methods; it can guess 78\% of ages within $\pm$2 years, 81\% of salaries within $\pm1000\$$, and almost all ages/salaries within $\pm$5 years and $\pm2000\$$, respectively, (ii) the accuracy of all methods increases as the acceptable age/salary range increases, and (iii) although our min-max based attack gives higher accuracy in comparison to random min-max (up to 35\% in age feature), no method can guess the target value very accurately just by using the minimum and maximum of the data.

\subsection{Performance Evaluation of our Protocols}
In order to measure the performance of \nsgx and \onesgx, we split the data among one active party and one passive party. We set the maximum depth of individual regression trees to 3, the samples' fraction for fitting the regression trees to 0.8, and  the learning rate to $0.3$. We used 128-bit AES encryption and 2048-bit Paillier encryption in \nsgx/\onesgx and \secureboost, respectively. We focus on measuring computation time and communication size (not including transmission time).

Figures~\ref{fig:performance}(c),(d) show the computation time and communication size of \nsgx, \onesgx, and \secureboost. As is clear from the figures, both our methods are 1.2-5.4$\times$ faster than \secureboost, while eliminating its leakage from split-finding. 
For instance, training a model with $10^5$ samples \secureboost takes 1697s while \nsgx and \onesgx take 314s and 323s, respectively. The main reason for the improvement in computation time is the use of AES and one-time pad instead of Paillier, which makes the cryptographic part of the computation much more efficient. Regarding communication size, we observe that \nsgx is the most efficient construction because it only transfers the passive parties' best local splits while \onesgx and \secureboost need to transfer all possible splits to the active party. Note that, since \onesgx uses symmetric encryption instead of Paillier, its communication size is less than \secureboost (we note that \secureboost cannot benefit from the classic ciphertext-packing technique to reduce communication, as each encrypted value needs to be accessed separately in order to compute all possible splits). According to our experiments, \nsgx and \onesgx need 32-49$\times$ and 5-10$\times$ less communication than \secureboost. E.g., for training a model with $10^5$ samples \secureboost needs to transfer a total of 307MB of data while \nsgx and \onesgx need only 9MB and 33MB, respectively.


{\small
\bibliographystyle{plain}
\bibliography{references}
}

\appendix
\section{\secureboost Routines}\label{sec:appendix}
For completeness, we include here the pseudocode of the \secureboost algorithms from~\cite{cheng2019secureboost}.
\begin{algorithm}
 \centering
    \caption{\secureboost: Aggregate Encrypted Gradient Statistics}\label{alg-secboost1}
    \begin{algorithmic}[1]
        \State \textbf{Input:} I, instance space of current node
        \State \textbf{Input:} d, feature dimension
        \State \textbf{Input:} $\{\langle g_i \rangle,\langle h_i \rangle\}_{i\in I}$ \Comment{$g_i$ and $h_i$ are gradients}
        \State \textbf{Output:} $G\in R^{d\times l}$, $H\in R^{d\times l}$
        \item[]
        \item[Passive Party] 
            \For{$k=0$ to $d$}
                \State $S_k=\{s_{k1},s_{k2},\dots,s_{kl}\}$ by percentiles on feature k
            \EndFor
            \For{$k=0$ to $d$}
                \State $G_{kv}=\sum_{i\in\{i|s_{k,v}\ge x_{i,k}>s_{k,v-1}\}}\langle g_i \rangle$ 
                \State $H_{kv}= \sum_{i\in\{i|s_{k,v}\ge x_{i,k}>s_{k,v-1}\}}\langle h_i \rangle$
            \EndFor
            \end{algorithmic}
\end{algorithm}
\begin{algorithm}
 \centering
    \caption{\secureboost: Split Finding}\label{alg-secboost2}
    \begin{algorithmic}[1]
        \State \textbf{Input:} I, instance space of current node
        \State \textbf{Input:} \{$G^i, H^i\}_{i=1}^m$ aggregated encrypted gradient statistics from m
        parties
        \State \textbf{Output:} partition current instance space according to the selected attribute's value
        \item[]
        \item[Active Party]
            \State $g\leftarrow \sum_{i\in I}g_i$, $h\leftarrow \sum_{i\in I}h_i$
            \State //enumerate all parties
            \For{$i=0$ to $m$}
                \State //enumerate all features
                \For{$k=0$ to $d_i$}  
                    \State $g_l=0, h_l=0$
                    \State //enumerate all threshold values
                    \For{$v=0$ to $l_k$}
                        \State get decrypted values $Dec_{sk}(G_{kv}^i)$ and $Dec_{sk}(H_{kv}^i)$
                        \State $g_l \leftarrow g_l + Dec_{sk}(G_{kv}^i)$
                        \State $h_l \leftarrow h_l + Dec_{sk}(H_{kv}^i)$ 
                        \State $g_r \leftarrow g-g_l$, $h_r \leftarrow h-h_l$
                        \State $score = max(score,\frac{g_l^2}{h_l+\lambda}+\frac{g_r^2}{h_r+\lambda}+\frac{g^2}{h+\lambda})$
                    \EndFor
                \EndFor
            \EndFor
            \State return $k_{opt}$ and $v_{opt}$ to the corresponding passive party $i$ when we obtain the max score
            \item[]
            \item[Passive Party $i$]
            \State determine the selected attribute's value according to $k_{opt}$ and $v_{opt}$ and partition current instance space
            \State record the selected attribute's value and return [record id, $I_L$] back to the active party
            \item[]
            \item[Active Party]
            \State split current node according to $I_L$ and associate current node with [party id,record id]
    \end{algorithmic}
\end{algorithm}
\end{document}